\documentclass[cits]{PoS}
\usepackage[numbers,sort&compress]{natbib}
\newcommand{\fesc}{\ifmmode{f_{\rm esc}}\else{$f_{\rm esc}$}\fi}
\newcommand{\fescs}{\ifmmode{f_{\rm esc}^\star}\else{$f_{\rm esc}^\star$}\fi}
\newcommand{\kms}{\ifmmode{{\rm km~s^{-1}}}\else{km~s$^{-1}$}\fi}
\newcommand{\fgas}{\ifmmode{{f_{\rm gas}}}\else{$f_{\rm gas}$}\fi}
\newcommand{\cubecm}{\ifmmode{{\;\rm cm^{-3}}}\else{cm$^{-3}$}\fi}
\newcommand{\ztwo}{\ifmmode{{\rm [Z_2/H]}}\else{[Z$_2$/H]}\fi}
\newcommand{\zthree}{\ifmmode{{\rm [Z_3/H]}}\else{[Z$_3$/H]}\fi}
\newcommand{\lsim}{\lower0.3em\hbox{$\,\buildrel <\over\sim\,$}}
\newcommand{\gsim}{\lower0.3em\hbox{$\,\buildrel >\over\sim\,$}}

\newcommand{\sfr}{M$_\odot$ yr$^{-1}$ Mpc$^{-3}$}

\newcommand{\Ms}{\ifmmode{\;\mathrm{M}_\odot}\else{$\mathrm{M}_\odot$}\fi}

\newcommand{\hh}{H$_2$}

\newcommand{\tvir}{\ifmmode{T_{\rm{vir}}}\else{$T_{\rm{vir}}$}\fi}
\newcommand{\mvir}{\ifmmode{M_{\rm{vir}}}\else{$M_{\rm{vir}}$}\fi}
\newcommand{\rvir}{\ifmmode{r_{\rm{vir}}}\else{$r_{\rm{vir}}$}\fi}

\newcommand{\lya}{Ly$\alpha$}
\newcommand{\jj}{\ifmmode{J_{21}}\else{$J_{21}$}\fi}
\newcommand{\flw}{\ifmmode{F_{LW}}\else{$F_{LW}$}\fi}
\newcommand{\kph}{\ifmmode{k_{\rm ph}}\else{$k_{\rm ph}$}\fi}

\newcommand{\zsun}{{\rm\,Z_\odot}}

\newcommand{\hii}{H {\sc ii}}

\def\eps@scaling{1.0}%
\newcommand\epsscale[1]{\gdef\eps@scaling{#1}}%
\newcommand\plotone[1]{%
 \centering 
 \leavevmode 
 \includegraphics[width={\eps@scaling\columnwidth}]{#1}%
}%
\newcommand\plottwo[2]{%
 \centering 
 \includegraphics[width={\eps@scaling\columnwidth}]{#1}%
 \hfil 
 \includegraphics[width={\eps@scaling\columnwidth}]{#2}%
}%

\title{First Light}

\ShortTitle{First Light}

\author{\speaker{John H. Wise}\thanks{The speaker appreciated the
    hospitality and efforts by the symposium organizers and the
    invitation to speak at this venue.  This review covers some of my
    own work, which could not have been accomplished without the help
    of my collaborators, Tom Abel, Marcelo A. Alvarez, Renyue Cen,
    Michael L. Norman, and Matthew J. Turk.}\\

  Center for Relativistic Astrophysics, Georgia Institute of
  Technology, 837 State Street, Atlanta, GA 30332, USA\\ 
  E-mail: \email{jwise@physics.gatech.edu}}

\abstract{The first stars in the universe are thought to be massive,
  forming in dark matter halos with masses around 10$^6$ solar masses.
  Recent simulations suggest that these metal-free (Population III)
  stars may form in binary or multiple systems.  Because of their high
  stellar masses and small host halos, their feedback ionizes the
  surrounding 3 kpc of intergalactic medium and drives the majority of
  the gas from the potential well.  The next generation of stars then
  must form in this gas-poor environment, creating the first galaxies
  that produce the majority of ionizing radiation during cosmic
  reionization.  I will review the latest developments in the field of
  Population III star formation and feedback and its impact on galaxy
  formation prior to reionization.  In particular, I will focus on the
  numerical simulations that have demonstrated this sequence of
  events, ultimately leading to cosmic reionization.}

\FullConference{Frank N. Bash Symposium New Horizons In Astronomy,\\
		October 9-11, 2011\\
		Austin, Texas}

\begin{document}

\section{Introduction}

The universe was a very dark place in the first tens of millions of
years before any luminous structure had formed.  This epoch is
sometimes referred to as the ``Dark Ages'', when dark matter (DM)
collapsed into bound objects but hosted no stars whatsoever.  These DM
halos collected a primordial mix of primarily hydrogen and helium in
their potential wells after they reached the cosmological Jeans mass
$M_J \sim 5 \times 10^3 [(1+z)/10]^{3/2}$ \cite{Barkana01}.  The
first DM halos to cool and collapse produced the first stars in the
universe that, in turn, produced the first metals to spark the
transition to galaxy formation.  Before reviewing the current status
of research on the first stars and galaxies, it is worthwhile to step
back, pose three simple but informative questions, and review
historical pieces of literature that addressed these questions.

The first question we can ask ourselves is: \textit{Why do all
  observed stars contain metals?}  \cite{Schwarzschild53} focused on
the Milky Way (MW) stellar population and observed (i) that the oldest
Population II stars were metal-poor, (ii) a high frequency of white
dwarfs, and (iii) a red excess in elliptical galaxies.  These points
led them to the conclusion that the ``original Population II contained
a large number of relatively massive stars''.  

The next question that naturally follows is: \textit{Without any
  metals, how does gas cool and condense to form stars?}
Metal-enriched gas cools mainly through \hh~formation on dust grains
and other fine-structure transitions in heavy elements.
\cite{McDowell61} recognized that \hh~can slowly form in the gas
phase through the following reactions:
\begin{eqnarray}
  \rm{H} + e^- &\rightarrow& \rm{H}^- + \gamma \nonumber\\
  \rm{H}^- + \rm{H} &\rightarrow& \rm{H}_2 + e^-  \nonumber
\end{eqnarray}
or less efficiently
\begin{eqnarray}
  \rm{H} + \rm{H}^+ &\rightarrow& \rm{H}_2^+ + \gamma  \nonumber\\
  \rm{H}_2^+ + \rm{H} &\rightarrow& \rm{H}_2 + \rm{H}^+ . \nonumber
\end{eqnarray}

\cite{Saslaw67} were the first to realize that \hh~formation in the
gas-phase was important in star formation in the early universe.  They
used these reactions to determine that \hh~cooling dominates the
collapse of a pre-galactic cloud at number densities $n > 10^4
\cubecm$.  These high density regions can cool to $\sim$300 K and
continue to collapse.  \cite{Peebles68} suggested that globular
clusters were the first bound objects in the universe with masses
$\sim$$5 \times 10^5 \Ms$.  In their calculations, they first compute
the properties of these objects from linear perturbation theory and
then follow the initial contraction of the cloud, including
\hh~cooling.  They find that molecular hydrogen cooling is indeed
efficient enough to drive a free-fall collapse, in which only a small
fraction of the total gas mass forms stars due to the inside-out
nature of the collapse.

The third pertinent question is: \textit{When and where do the first
  stars form?}  Applying the properties of cooling primordial gas to
the cold dark matter model, \cite{Couchman86} determined that the
first objects to cool and collapse due to \hh~formation would be
hosted in DM halos with masses $\sim 10^6 \Ms$ at $z = 20-30$.  A
decade later, \cite{Tegmark97} used a detailed chemical model to
follow the formation of \hh~in virialized objects, starting from
recombination.  They found that the redshift-dependent minimum DM halo
mass to host \hh~cooling, rising from $5 \times 10^3 \Ms$ at $z \sim
100$ to $10^6 \Ms$ at $z \sim 15$.

These early analytical works provided the theoretical basis for
current and recent simulations that focus on the formation of the
first stars and galaxies in the universe.  Here I review the progress
that the field has made in the past decade.

\section{Population III star formation}

Following a cosmological gaseous collapse over many orders of
magnitude in length makes such a simulation technically difficult, but
with improvements in algorithms and physical models, several groups
have been able to make substantial progress.  In the late 1990's and
early 2000's, two independent groups used three-dimensional
simulations to study Population III star formation.  Using smoothed
particle hydrodynamics (SPH) simulations of isolated and virialized
dark matter halos with masses $2 \times 10^6 \Ms$ at $z = 30$, one
group found that the object cooled to 300~K through \hh~formation,
using the chemical network of \cite{Haiman96}, and fragmented into a
filamentary structure with a Jeans mass of $10^3 \Ms$ \cite{Bromm99,
  Bromm02_P3}.  In the second paper, they followed the collapse to
higher gas densities of $10^8 \cubecm$ and studied the continued
formation of dense clumps with the same $10^3 \Ms$ characteristic
mass.  The other group used cosmological adaptive mesh refinement
(AMR) simulations to focus on the formation of a molecular cloud
hosted by a $7 \times 10^5 \Ms$ DM halo \cite{Abel98, Abel00, ABN02}.
In each successive paper, the \hh~chemistry model \cite{Abel97,
  Anninos97} was improved to include more processes, such as the
three-body \hh~formation process, to follow the central collapse to
gas densities up to $3 \times 10^{13} \cubecm$.  Both groups came to
the conclusion that further fragmentation was suppressed because of a
lack of cooling below 300~K and that Population III stars were
\textit{very massive} in the range $30-300 \Ms$.

These simulations only represented a limited sample of collapses and
could not provide much insight to the Population III initial mass
function (IMF).  Twelve additional AMR simulations were conducted to
look at any variations in primordial gas collapses \cite{OShea07a}.
They found that the collapses occurred in DM halos with masses in the
range $1.5-7 \times 10^5 \Ms$ with the scatter caused by differing
halo formation histories.  The mass accretion rate onto the central
molecular cloud was higher at $10^{-4} \Ms \; \mathrm{yr}^{-1}$ at $z
\sim 30$, and it can increase by two orders of magnitude at $z \sim
20$ in some halos, agreeing with \cite{ABN02}.

Following the collapse to densities higher than $10^{13} \cubecm$
required the inclusion of collisionally induced emission, chemical
heating from \hh~formation, and gas opacity above $10^{18} \cubecm$.
\cite{Yoshida08} found with SPH simulations that the initial
collapsing region did not fragment as it condensed to protostellar
densities $n \sim 10^{21} \cubecm$, forming a protostellar shock in
the process.  The inner 10 \Ms~had an accretion rate varying between
0.01 and 0.1 \Ms~yr$^{-1}$, possibly growing to 10 \Ms~within 1000 yr.

In the past few years, multiple groups have been focusing on the
subsequent growth of these protostars over several dynamical times,
improving upon the earlier works that stopped at the first collapse.
This has proved to be challenging because of the ever-decreasing
Courant factors at higher densities.  One workaround is the creation
of ``sink particles'' that accrete nearby gravitationally-bound gas,
allowing the simulation to progress past the first collapse; however,
one loses all hydrodynamical information above some density threshold.
In one out of five realizations in AMR calculations without sink
particles, \cite{Turk09} found that the collapsing core fragmented at
a density of $10^{11} \cubecm$ into two clumps that are separated by
800 AU with 100 \Ms~of gas within a sphere with radius twice their
separation.  At the same time, \cite{Stacy10_Binary} also found that
disk instabilities causes fragmentation into a binary system with a
40~\Ms~and 10~\Ms.  This was later confirmed by simulations of a
collapse of an isolated Jeans-unstable primordial gas cloud that
fragmented into many multiple systems with some very tightly bound to
separations less than an AU \cite{Clark11_Frag}.  Utilizing a new
moving mesh code, \cite{Greif11_P3Cluster} studied the collapse in
five different primordial DM halos, and they evolved them for 1000 yr
after the first protostar forms.  By evolving these protostars
further, they included the effects of protostellar radiative feedback
in the infrared in the optically-thin limit.  In all cases, the
molecular cloud fragments into $\sim 10$ sink particles, some of which
later merge to form more massive protostars.  The mass function from
these simulations is relatively flat, i.e. a top-heavy IMF.

After the protostar has reached $\sim 10 \Ms$, radiative feedback from
ionizing radiation will begin to suppress further accretion.  Only
recently has this been incorporated into numerical simulations of
Population III star formation.  Starting from initial conditions
extracted from a cosmological simulation \cite{Yoshida08},
two-dimensional axi-symmetric radiation hydrodynamical simulations
showed that an accretion disk forms around a new protostar with the
ionizing radiation preferentially escaping through the polar regions
\cite{Hosokawa11}.  The disk itself is slowly photo-evaporated,
halting accretion after 70,000 yr.  At this point, the final mass of
the Population III star is 43 \Ms.  Without any radiative feedback,
the protostar would have continued to grow to $\sim 100 \Ms$.  In a
cosmological setting, \cite{Stacy11} found a binary system still
forms in the presence of radiative feedback.  Without feedback, the
primary star grows to 28~\Ms~over 5,000 yr.  With feedback, the
primary and secondary stars only grow to 19 and 10~\Ms, respectively.
An extrapolation of the mass accretion history shows that both stellar
masses will asymptote to 30~\Ms, creating an equal-mass binary.  Once
the stars have entered the main sequence, they will start to ionize
and heat their cosmic neighborhood, which I will review next.

\section{Population III radiative feedback}

Some of the first calculations of the growth and overlap of
cosmological \hii~regions originating from quasars concluded that they
could not fully account for reionization.  Other radiation sources
must have contributed to the photon budget \cite{Shapiro86}.  Later
with $z>4$ galaxy observations, it was clear that low-luminosity
galaxies were the primary source of ionizing photons \cite[e.g.][]
{Bouwens04, Fan06, Thompson07}.  However, Population III preceded
galaxy formation, and they were the first sources of ionizing
radiation, starting cosmic reionization.  They are thought to have a
top-heavy IMF, as discussed in the previous section, and
zero-metallicity stellar models were constructed to estimate their
luminosities, lifetimes, and spectra as a function of mass
\cite{Tumlinson00, Bromm01_Sp, Schaerer02}.  One key feature is
mass-independent surface temperature of $10^5$ K above 40~\Ms, caused
by a lack of opacity from metal lines.  Thus, Population III are
copious producers of ionizing photons with an average photon energy
$\sim 30$ eV and also \hh~dissociating radiation, which is $<13.6$ eV
where the neutral universe is optically-thin.  Because the formation
of Population III stars is primarily dependent on \hh~formation, any
\hh~dissociating radiation can suppress Population III star formation
from large distances \cite{Dekel87, Machacek01, Johnson07,
  Wise07_UVB, OShea08}.

\subsection{\hii~regions from Population III stars}

Combining the main sequence properties of Population III stars and the
endpoints of cosmological simulations, one-dimensional radiation
hydrodynamics simulations followed the growth of an \hii~region from
Population III stars with masses ranging from 25 to 500
\Ms~\cite{Kitayama04, Whalen04}.  The ionization front drives a
30~\kms~shock wave.  Because the escape velocity of $10^6 \Ms$ halos
is only $\sim$ 3 \kms, approximately 90\% of the gas is expelled from
the DM halo, leaving behind a warm ($T \sim 3 \times 10^4$ K) and
diffuse ($\rho \sim 0.1 \cubecm$) medium.  At the end of the star's
lifetime, a 100~\Ms~star creates an \hii~region with a radius $\sim 3$
kpc.

\begin{figure}
\includegraphics[width=\textwidth]{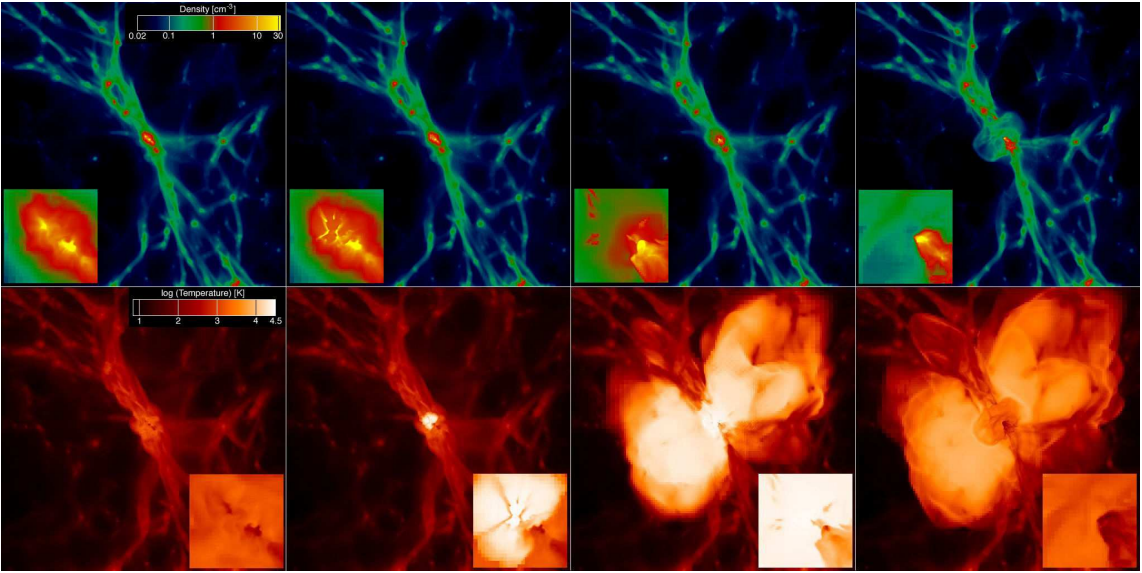}
\caption{\label{fig:p3rt} The formation of a \hii~region from a
  Population III star, shown with projections of gas density (top) and
  temperature (bottom) of a $\sim 3$ proper kpc region, centered on
  the first star at $z = 20$.  From left to right, the depicted times
  correspond to 0, 1, 2.7, and 8 Myr after the star formed.  The
  insets correspond to the same times, have the same color scale, and
  show the central 150 pc.  From \cite{Abel07}.}
\end{figure}

Shortly afterward, it became feasible to include radiative transfer in
cosmological simulations, either through moment methods or ray tracing.
In three dimensions, it is possible to investigate the ionization of a
clumpy and inhomogeneous medium and any ionization front instabilities
\cite{Whalen11} that might arise.  \cite{Alvarez06_IFT} found that
between 70\% and 90\% of the ionizing photons escaped into the IGM,
using ionization front tracking and an approximate method to calculate
the thermodynamic state behind the front.  This calculation also
showed that some nearby halos are not fully photo-evaporated, leaving
behind a neutral core.  Furthermore, nearby filamentary structure is
slower to ionize, and the ionization front grows faster in the voids,
creating a ``butterfly'' shape \cite{Abel99_RT}.  The first
three-dimensional radiation hydrodynamics simulations uncovered
multi-fold complexities that were not seen in previous simulations,
such as cometary structures and elephant trunks seen in nearby star
forming regions \cite{Abel07}.  Figure \ref{fig:p3rt} shows the
growth of the \hii~region emanating from the host minihalo.  The
density structures in the nearby filaments were largely unaffected by
the radiation because they are self-shielded.  The 30~\kms~shock wave
collects $10^5 \Ms$ of gas into a shell over the lifetime of the star,
which is Jeans stable and is dispersed after the star's death.

As the \hii~region grows up to 3~kpc in radius, nearby halos become
engulfed in a sea of ionized gas.  Because free electrons are the
catalyst for \hh~formation, a boosted electron fraction promotes more
efficient cooling; furthermore, HD cooling becomes relevant in the
collapse of these halos in relic \hii~regions \cite{OShea05,
  Yoshida07_HII, Yoshida07, McGreer08}.  Instead being limited to a
300~K temperature floor, this gas cools to $\sim 50$ K, resulting in a
Jeans mass a factor of $\sim 5/2$ lower.  Thus, it is expected that
these Population III stars will have a lower characteristic mass in
the approximate range of 5--60~\Ms.  After this discovery, it was felt
that these two different population needed to be separated, where
metal-free stars forming in an unaffected region are termed
``Population III.1'', and metal-free stars forming in ionized gas
were coined ``Population III.2'' \cite{Norman08}.

\subsection{Contribution to reionization}

\begin{figure}
\includegraphics[width=\textwidth]{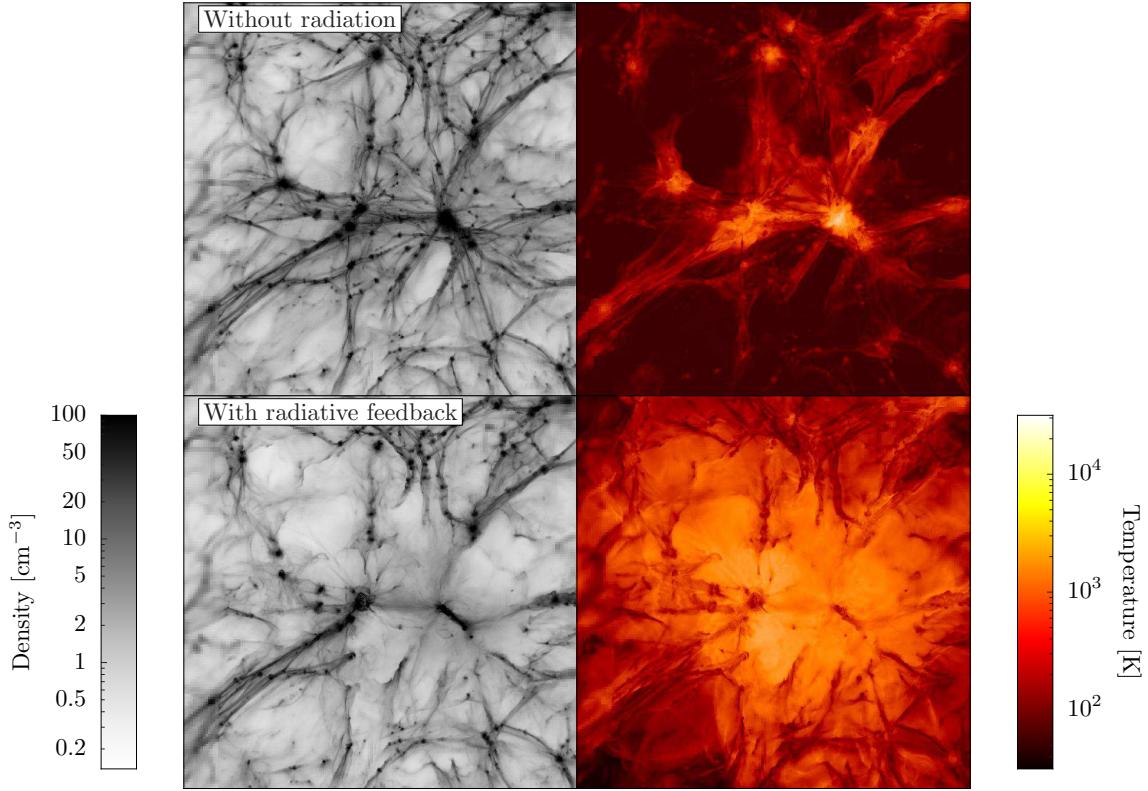}
\caption{\label{fig:wise_reion} The effects of radiative feedback from
  the first stars, shown in projections of gas density (left) and
  temperature (right) in a field of view of 8.5 proper kpc in a region
  heated and ionized by tens of Population III stars at $z = 16$.
  Notice how most of the nearby substructures are photo-evaporated.
  From \cite{Wise08_Reion}.}
\end{figure}

The \hii~regions from metal-free stars are much larger than
present-day \hii~regions and can have a sizable impact on the
reionization history.  \cite{Wise08_Reion} found that Population III
stars can ionize up to 25\% of the local IGM in a biased region,
surrounding a rare $3-\sigma$ overdensity.  Because Population III
stars are short lived ($\sim 3$ Myr), the \hii~regions are fully
ionized only for a short time and then quickly recombines over the
next $\sim 50$ Myr.  Within a local cosmological region, there are
many relic \hii~regions but only a handful of active \hii~regions with
$T > 10^4$ K.  Once the \hii~regions start to overlap, each star can
ionize a larger volume as the neutral hydrogen column density
decreases.  At the end of the simulation, one in ten ionizing photons
results in a sustained ionization in the intergalactic medium (IGM).
In addition to ionizing the IGM, the photo-heating of the host halo
and IGM delays further local star formation by smoothing out gas
overdensities in nearby minihalos and IGM, which is depicted in Figure
\ref{fig:wise_reion}.  Reducing the IGM clumpiness reduces the
recombination rate, which is measured by the clumping factor $C =
\langle \rho \rangle^2 / \langle \rho^2 \rangle$, by 50\%
\cite{Wise08_Reion}.

This simulation only considered a small region (1 comoving Mpc$^3$)
and cannot make predictions for global reionization history.  To
address cosmic reionization, simulations with sizes $\sim 100$ Mpc are
necessary.  Here the small-scale clumpiness cannot be resolved, and
clumping factor plays a key role in subgrid models.  In general,
\hii~regions from Population III stars generate more small-scale
power, and at late times, they are quickly overrun by nearby
\hii~regions produced by larger galaxies \cite{Iliev07}.  In
addition, Population III \hii~regions start reionization earlier and
prolongs reionization \cite{Trac07}.

\section{Supernovae from Population III}

Massive metal-free stars can end their lives in a unique type of
supernova, a pair-instability SN \cite[e.g.][]{Barkat67, Bond84,
  Heger02}.  Non-rotating models find that this occurs in a mass range
between 140 and 260~\Ms, where nearly all of the helium core with mass
$M_{\rm He} \approx 13/24 (M_\star - 20 \Ms)$ is converted into metals
in an explosion of $10^{51} - 10^{53}$ erg.  The ejecta can be an
order of magnitude greater than typical Type II SNe \cite{Woosley95}
and hypernovae \cite{Nomoto06}!  The chemical abundance patterns are
much different than those in typical explosions with the carbon,
calcium, and magnesium yields independent of mass.  These
pair-instability SNe are one possible cause for carbon-enhanced damped
\lya~absorbers \cite[e.g.][]{Penprase10, Cooke11}.

These very energetic SNe can exceed the binding energy of halos with
masses $M \lsim 10^7 \Ms$.  \cite{Bromm03} investigated two explosion
energies, $10^{51}$ and $10^{53}$ erg, in a cosmological halo with $M
\sim 10^6 \Ms$, neglecting any radiative feedback.  Nevertheless, they
found that over 90\% of the gas was expelled into the IGM, and metals
propagate to distances of $\sim 1$ kpc after 3--5 Myr.  They argued
that pair-instability SNe could have resulted in a nearly uniform
metallicity floor in the IGM of $\sim 10^{-4} Z_\odot$ at high
redshifts.  Subsequent works built upon this idea of a IGM metallicity
floor with various techniques: (i) volume-averaged semi-analytic
models \cite{Scannapieco03, Yoshida04, Furlanetto05_Reion}, (ii)
models using hierarchical merger trees \cite{Tumlinson06,
  Salvadori07, Komiya10}, (iii) post-processing of cosmological
simulations with blastwave models \cite{Karlsson08, Trenti09}, and
(iv) direct numerical simulations with stellar feedback
\cite{Tornatore07, Ricotti08, Maio11_Enrich, Wise12_Galaxy}.

Because blastwaves do not penetrate overdensities as efficiently as a
rarefied medium, the voids will be preferentially enriched
\cite{Cen08}.  This raises the following questions.  Will the first
galaxies have a similar metallicity as the IGM?  How much metal mixing
occurs in the first galaxies as they accrete material?  The complex
interplay between radiative and supernova feedback, cosmological
accretion, and hydrodynamics are best captured by numerical
simulations.  Two groups \cite{Wise08_Gal, Greif10} showed that the
enrichment from pair-instability SNe resulted in a nearly uniform
metallicity in a $10^8 \Ms$ halo at $z \sim 10-15$.  These types of
halos can efficiently cool through atomic hydrogen cooling, and the
halo will form a substantial amount of stars for the first time.  Both
groups find that the metals are well-mixed in the galaxy because of
turbulence generated during virialization \cite{Wise07, Greif08} to a
metallicity $Z/Z_\odot = 10^{-3} - 10^{-4}$.  In these simulations,
about 60\% of the metals from SNe are reincorporated into the halo,
whereas the remaining fraction stays in the IGM.  In the end,
Population III star formation is ultimately halted by the enrichment
of the minihalos from nearby or previously hosted supernovae (SNe),
marking the transition to galaxy formation.

\section{High-redshift dwarf galaxies}

\begin{figure}
\includegraphics[width=\textwidth]{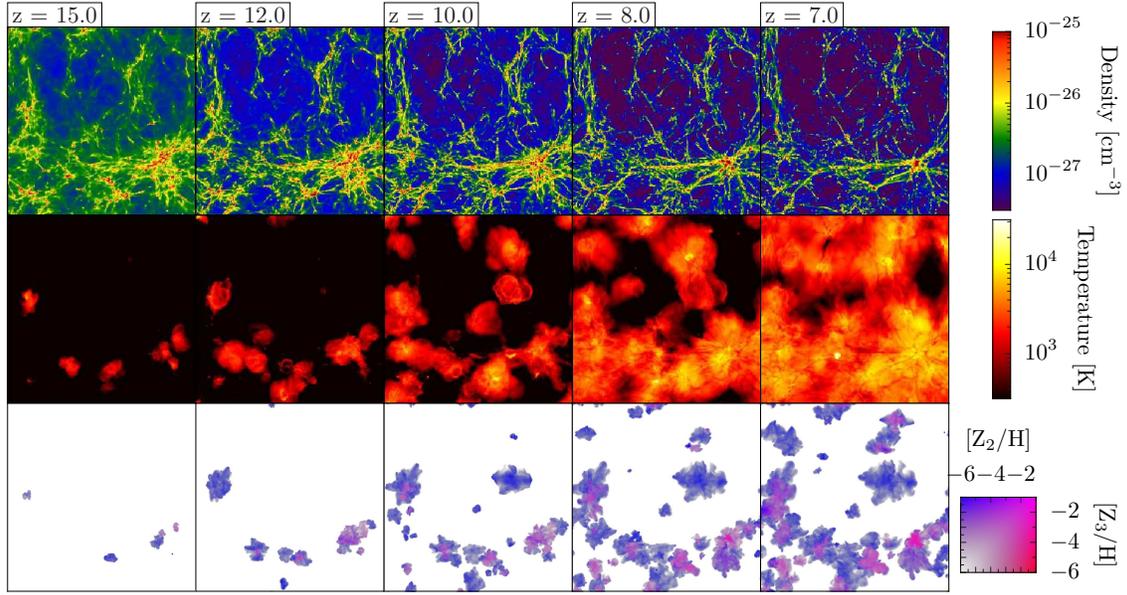}
\caption{\label{fig:galevo} Evolution of the entire simulation volume
  ($L_{\rm box} = 1$ Mpc) at redshifts 15, 12, 10, 8, and 7 (left to
  right) that follows the formation of 38 dwarf galaxies and over 300
  Population III stars.  Pictured here are the density-weighted
  projections of density (top), temperature (middle), and metallicity
  (bottom).  Note how the stellar radiative feedback from low-mass
  galaxies reionize the majority of the volume.  The metallicity
  projections are a composite image of metals originating from Pop II
  (red) and III (blue) stars with magenta indicating a mixture of the
  two.  From \cite{Wise12_Galaxy}.}
\end{figure}

\begin{figure}
\includegraphics[width=0.475\textwidth]{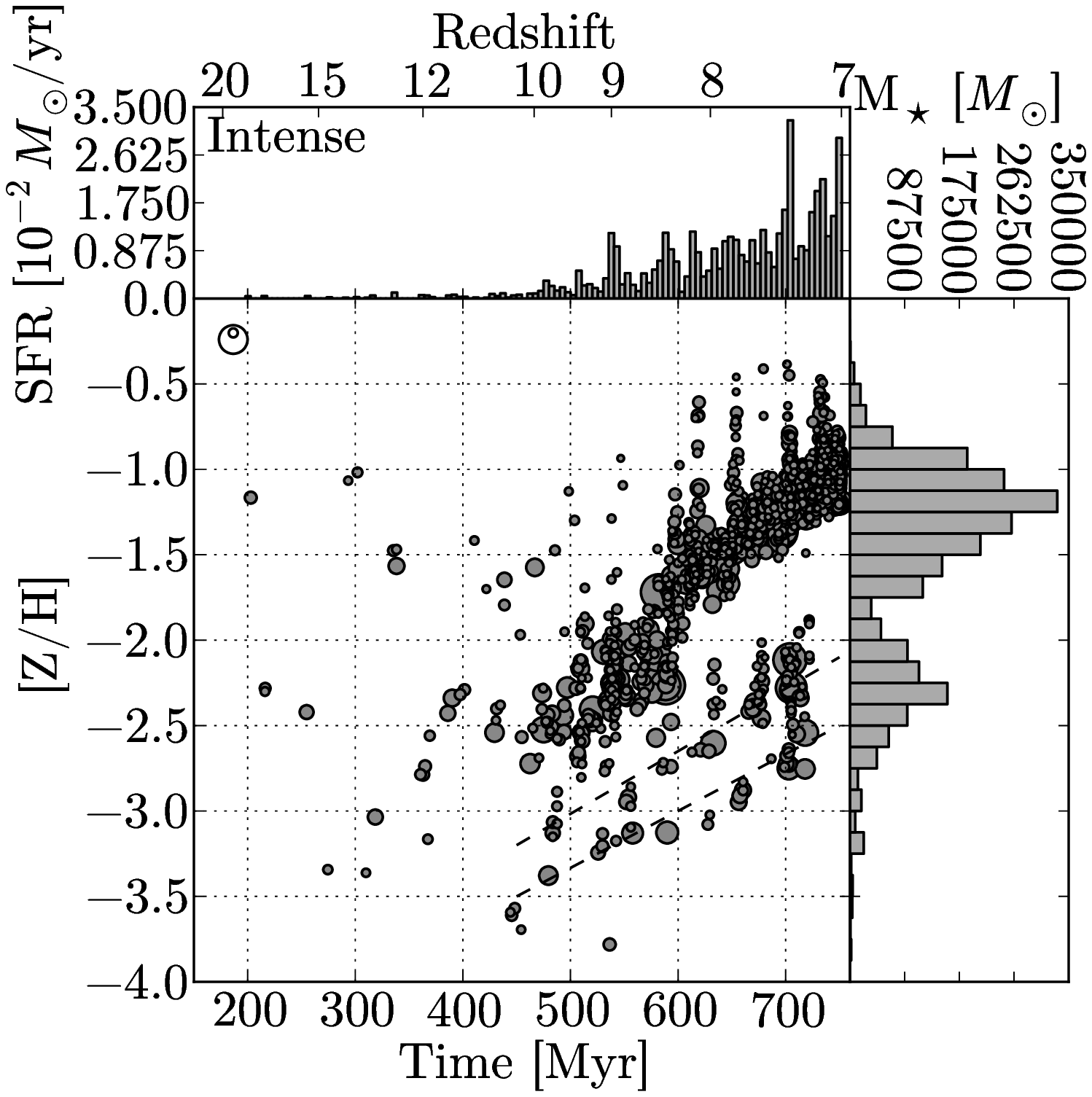}
\hfill
\includegraphics[width=0.475\textwidth]{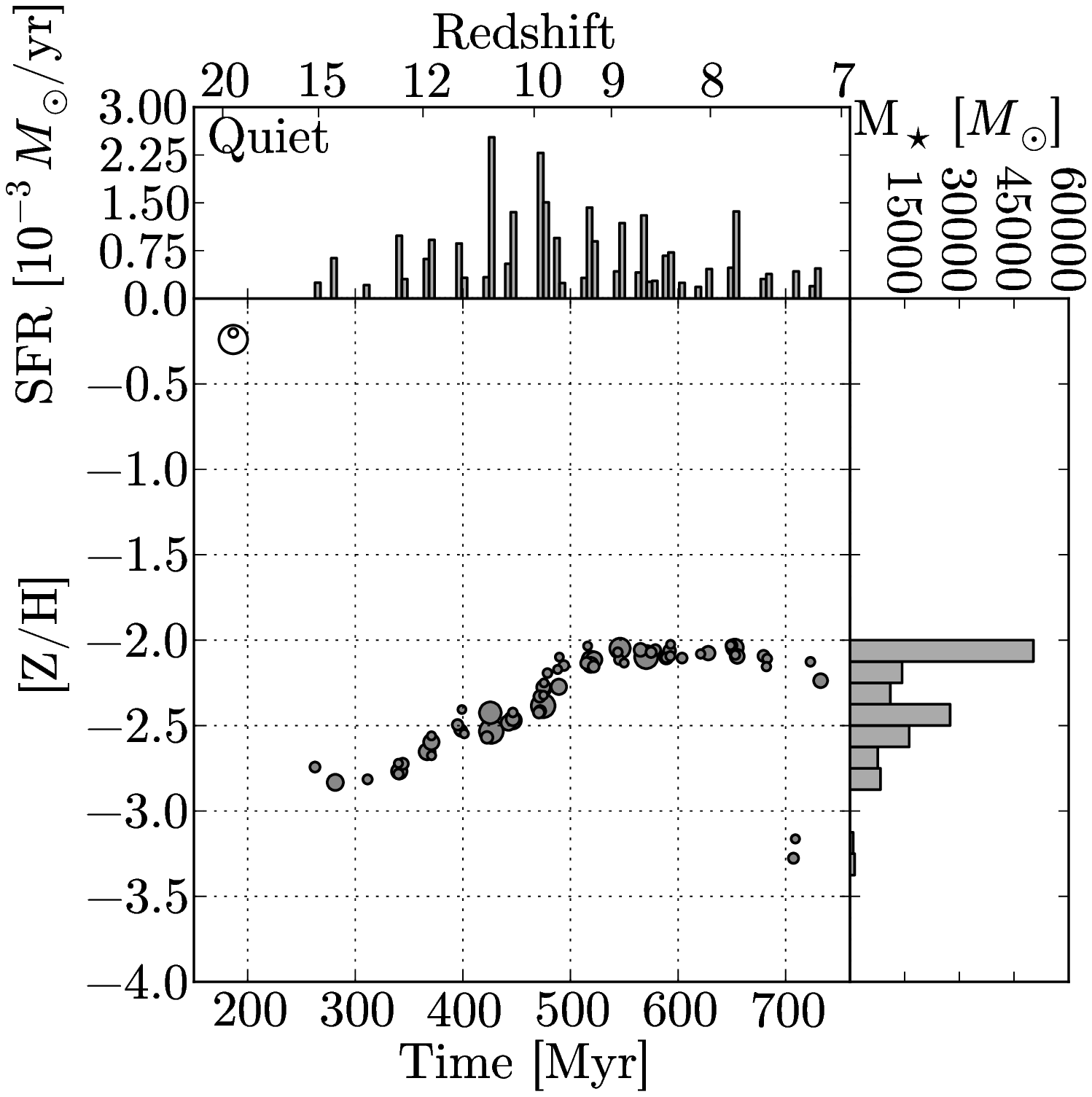}
\caption{\label{fig:sfh} The scatter plots show the metal-enriched
  (Pop II) star formation history of a $10^9 \Ms$ (left) and a $10^8
  \Ms$ (right) halos as a function of total metallicity, i.e. the sum
  of metal ejecta from both Pop II and Pop III SNe, at $z=7$.  Each
  circle represents a star cluster, whose area is proportional to its
  mass.  The open circles in the upper right represent $10^3$ and
  $10^4$ \Ms~star clusters.  The upper histogram shows the SFR.  The
  right histogram depicts the stellar metallicity distribution.  The
  larger halo shows a large spread in metallicity at $z>10$ because
  these stars formed in progenitor halos that were enriched by
  different SN explosions.  At $z<10$, the majority of stellar
  metallicities increase as the halo is self-enriched.  The spikes in
  metallicity at $t$ = 620, 650, and 700 Myr show induced star
  formation with enhanced metallicities in SN remnant shells.  The
  dashed lines in the left panel guide the eye to two stellar
  populations that were formed in two satellite halos, merging at
  $z=7.5$.  The smaller halo evolves in relative isolation and
  steadily increases its metallicity to [Z/H] $\sim$ --2 until there
  is an equilibrium between \textit{in-situ} star formation and
  metal-poor inflows from filaments. From \cite{Wise12_Galaxy}.}
\end{figure}

The first galaxies are generally defined as halos that can undergo
atomic line cooling, are metal-enriched, and can host sustained star
formation \cite{Bromm11}.  Here I present some of the highlights of
our latest numerical work on the formation of the first galaxies
\cite{Wise12_Galaxy}.  These radiation hydrodynamics AMR simulations
tracked the formation and feedback of over 300 Population III stars
and the buildup of 38 low-mass galaxies in a 1 comoving Mpc$^3$ volume
until $z=7$.  The cosmic Population III star formation rate (SFR) is
nearly constant at $3 \times 10^{-5}$ \sfr~from $z=15$ to $z=7$.  The
largest galaxy has a final total and stellar mass of $1.0 \times 10^9
\Ms$ and $2.1 \times 10^6 \Ms$, respectively.  Galaxies above $10^8
\Ms$ generally have a mass-to-light ratio between 5 and 30, whereas
the very low-mass galaxies have mass-to-light ratios between 100 and
3000 because of their inability to efficiently form stars.  

The evolution of the density, temperature, and metallicity of the
entire volume is shown in Figure \ref{fig:galevo}.  At $z=7$, 76\% of
the volume is ionized, and 6.5\% (1.9\%) of the mass (volume) is
enriched above $10^{-3} \zsun$.  We focused on the buildup of the
largest galaxy and an isolated dwarf galaxy with a total mass of $10^8
\Ms$.  Figure \ref{fig:sfh} shows the metallicity of the star
formation history and metallicity distribution functions in both
halos.  The mass resolution of this simulation captures the formation
of all star-forming minihalos with $M > 10^5 \Ms$.

The smaller galaxy experiences rapid mass accretion until $z \sim 12$
and afterward it evolves in relative isolation.  It begins forming
metal-enriched stars after a nearby pair-instability SN enriches a
nearby halo to $\sim 10^3 \zsun$.  This may be a peculiar case at high
redshift, where a halo is enriched from a neighboring halo and does
not form any Population III stars itself.  It begins to form stars in
a bursts at a rate of $5 \times 10^{-4}$ \sfr, peaking at $2 \times
10^{-3}$ \sfr~at $z=10$.  The galaxy is self-enriched by these stars,
gradually increasing from $10^{-3} \zsun$ to $10^{-2} \zsun$ by
$z=10$.  Afterward there is an equilibrium between metal-rich outflows
and metal-poor accretion from the filaments, illustrated by the
plateau in stellar metallicities in Figure \ref{fig:sfh}.

The larger galaxy forms in a biased region of 50 comoving kpc on a
side with $\sim 25$ halos with $M \sim 10^6 \Ms$ at $z=10$.  About
half of these halos form Population III stars with a third producing
pair-instability SNe, enriching the region to $10^{-3} \zsun$, the
metallicity floor that has been extensively studied in previous works.
However, the metal-rich ejecta does not fully escape from the biased
region, and most of it falls back into the galaxies or surrounding
IGM, leaving the voids pristine.  After $z=10$, these $\sim 25$ halos
hierarchically merge to form a $10^9 \Ms$ halo at $z=7$ with two major
mergers at $z=10$ and $z=7.9$.  At late times, this galaxy grows
mainly through mergers with halos above the filtering mass
\cite{Gnedin98, Gnedin00, Wise08_Reion}, i.e. gas-rich halos that are
not photo-evaporated, and the gas fraction increases from 0.08 to 0.15
over the last 200 Myr of the simulation.  The left panel of Figure
\ref{fig:sfh} shows a large scatter in metallicity at early times,
which is caused by inhomogeneous metal enrichment of its progenitors.
Once it hosts sustained star formation after $z=10$, the metallicity
trends upwards as the stars enriches its host galaxy.  In contrast
with the smaller halo, the larger galaxy undergoes a few mergers with
halos with an established stellar population.  This creates a
superposition of age-metallicity tracks in the star formation history.

This simulation of the early stages of galaxy formation only covered a
handful of galaxies and did not explore the differing galaxy
populations.  However, it has given us a clear picture of the inner
workings of these galaxies and the important physical processes
involved in shaping the first galaxies and their connections to the
first stars.  We hope to improve on this work to survey a larger
galaxy population and focus on larger galaxies that the \textit{Hubble
  Space Telescope} has already observed and the \textit{James Webb
  Space Telescope} will observe at $z > 6$.

\section{Summary}

I have provided a brief review of the formation of the first stars and
their radiative, chemical, and mechanical feedback that affects
subsequent structure and galaxy formation.  Over the past decade, many
groups have used numerical simulations to study these astrophysics
events in the first billion years of the universe.  Currently, the
general consensus is that Population III stars are still very massive
with a characteristic mass of tens \Ms~with an unknown fraction in
binaries.  The prospect of Population III binaries is exciting, and
their impact on the universe prior to reionization, such as
pre-ionization from X-rays, will be addressed in future studies.  To
summarize, the radiation from Population III expels most of the gas
from the host halos, creating gas-poor halos that cannot form stars
for 10--50 Myr.  The SNe from the first stars enriches the first
galaxies to a nearly uniform $\sim 10^{-3} \zsun$, and ultimately
leads to the demise of this unique population of stars.  The gas
depletion, IGM pre-heating, and chemical enrichment all have a lasting
impact on the formation of the first galaxies.  Hopefully we can
utilize these imprints to disentangle Population III stellar
properties from the most distant galaxies in the universe.

\bibliography{jwise}

\end{document}